\documentclass[12pt,a4paper,portuguese]{article}
\usepackage{amssymb}

\usepackage{graphicx}
\usepackage{amsmath}
\usepackage{babel}
\usepackage{a4}

\input{tcilatex}

\begin{document}

\begin{center}
{\Large A Inven\c{c}\~{a}o do Conceito de Quantum de Energia segundo Planck}

\medskip

Nelson Studart

\smallskip

\textit{Departamento de F\'{\i}sica, Universidade Federal de S\~{a}o Carlos,
13565-905, S\~{a}o Carlos, S\~{a}o Paulo}

\bigskip

Recebido em 22 de Novembro de 2000. Aceito em 29 de Dezembro de 2000.

\medskip
\end{center}

{\small H\'{a} cem anos, a id\'{e}ia do quantum provocou uma revolu\c{c}\~{a}%
o na ci\^{e}ncia e a busca de uma nova base conceitual para a toda a f\'{\i}%
sica, como enfatizou Einstein. Neste artigo, discuto os aspectos essenciais
dos trabalhos de Planck de 1900 sobre a radia\c{c}\~{a}o do corpo negro e a
hip\'{o}tese da quantiza\c{c}\~{a}o da energia.}

{\small \bigskip }

{\small A hundred years ago, the quantum concept provoked a revolution in
science and the search of a new conceptual basis for whole physics, as
emphasized by Einstein. In this paper, I discuss the essential features of
Planck's works in 1900 on the blackbody radiation and the hypothesis of
energy quantization. }

\section{Introdu\c{c}\~{a}o}

\begin{quotation}
{\small Trope\c{c}avas nos astros desastrada}

{\small Quase n\~{a}o t\'{\i}nhamos livros em casa}

{\small E a cidade n\~{a}o tinha livraria}

{\small Mas os livros que em nossa vida entraram}

{\small S\~{a}o como a radia\c{c}\~{a}o do corpo negro}

{\small Apontando pra a expans\~{a}o do Universo}

{\small Porque a frase, o conceito, o enredo, o verso}

{\small (E, sem d\'{u}vida, sobretudo o verso)}

{\small \'{E} o que pode lan\c{c}ar mundos no mundo.}

{\small \smallskip }

{\small Caetano Veloso em Livros (1a. estrofe)}

\smallskip
\end{quotation}

Em 14 de dezembro de 1900, Max Karl Ernst Ludwig Planck (1858-1957)
apresentou, em uma reuni\~{a}o da Sociedade Alem\~{a} de F\'{\i}sica, o
trabalho intitulado \textit{Sobre a Teoria da Lei de Distribui\c{c}\~{a}o de
Energia no Espectro Normal}\cite{planck00-14}, em que introduziu o conceito
de quantiza\c{c}\~{a}o da energia e deu origem a uma das revolu\c{c}\~{o}es
da f\'{\i}sica no s\'{e}culo XX. Os versos de Caetano Veloso, do seu disco 
\textit{Livro}\footnote{%
Agraciado este ano com o \textit{Grammy}, a mais importante premia\c{c}\~{a}%
o musical dos Estados Unidos, na categoria \textit{World Music.}}, torna o
tema bastante atual para o grande p\'{u}blico ao enfocar a radia\c{c}\~{a}o c%
\'{o}smica de fundo, o f\'{o}ssil remanescente do Universo primordial\textit{%
\ }e a principal evid\^{e}ncia da exist\^{e}ncia do \textit{big bang} e
expans\~{a}o do Universo.

As origens da teoria qu\^{a}ntica, em particular o significado e a repercuss%
\~{a}o da hip\'{o}tese de Planck, t\^{e}m sido analisadas em v\'{a}rios
artigos e livros com o devido rigor hist\'{o}rico. Dentre eles, gostaria de
destacar os de Leon Rosenfeld\cite{rosenfeld36}, Martin Klein \cite%
{klein62,klein63,klein66,klein77}, Hans Kangro\cite{kangro72,kangro76}, Max
Jammer\cite{jammer89}, e Mehra e Rechenberg\cite{mehra82} que relevam
sobremaneira o papel de Planck como inventor da teoria qu\^{a}ntica.

Mais do que celebrar este importante acontecimento, pretendo neste artigo
discutir as origens da teoria qu\^{a}ntica de um modo did\'{a}tico $-$ pelo
menos na concep\c{c}\~{a}o do autor $-$ com a esperan\c{c}a de que este
texto possa ser usado em disciplinas de F\'{\i}sica Moderna e Mec\^{a}nica Qu%
\^{a}ntica, com \^{e}nfase em seus aspectos hist\'{o}ricos, bem como na
disciplina de Evolu\c{c}\~{a}o de Conceitos da F\'{\i}sica, ou mesmo, de Hist%
\'{o}ria da F\'{\i}sica.\footnote{%
Uma introdu\c{c}\~{a}o bastante simples aos conceitos da f\'{\i}sica qu\^{a}%
ntica numa abordagem quase-hist\'{o}rica pode ser encontrada na Ref. \cite%
{cropper70}.}\ 

A inven\c{c}\~{a}o do quantum de energia \'{e} um dos muitos exemplos na hist%
\'{o}ria da ci\^{e}ncia que revela que ``conceitos cient\'{\i}ficos s\~{a}o
criados por a\c{c}\~{o}es da imagina\c{c}\~{a}o e intelig\^{e}ncia humanas e
n\~{a}o s\~{a}o como objetos que s\~{a}o `descobertos' como entidades que j%
\'{a} existem''.\cite{arons97} Como enfatiza Arons, alertar os estudantes
para este fato \'{e} uma tarefa revestida de car\'{a}ter pedag\'{o}gico
inestim\'{a}vel.

Uma outra caracter\'{\i}stica relevante desta inven\c{c}\~{a}o \'{e} a
demonstra\c{c}\~{a}o cabal da efici\^{e}ncia da intera\c{c}\~{a}o entre o
pesquisador te\'{o}rico e o experimental provido de uma infra-estrutura
laboratorial adequada. A liga\c{c}\~{a}o estreita entre Planck e seus
colegas Heinrich Rubens (1865-1922) e Ferdinand Kurlbaum (1857-1927), do 
\textit{Physicalisch-Technische Reichsanstalt} $-$ o mais importante laborat%
\'{o}rio alem\~{a}o f\'{\i}sico-t\'{e}cnico, centro de refer\^{e}ncia de
pesos e medidas e precursor do atual Laborat\'{o}rio Nacional de F\'{\i}sica
e Tecnologia da Alemanha $-$ foi fundamental para que a hip\'{o}tese da
quantiza\c{c}\~{a}o da energia fosse formulada. Como veremos, foi a constata%
\c{c}\~{a}o, por Rubens, de que seus dados experimentais se ajustavam muito
bem \`{a} f\'{o}rmula de Planck para a distribui\c{c}\~{a}o espectral da
radia\c{c}\~{a}o t\'{e}rmica, obtida atrav\'{e}s da interpola\c{c}\~{a}o
entre dois limites de freq\"{u}\^{e}ncias, que levou Planck a procurar as raz%
\~{o}es b\'{a}sicas para a obten\c{c}\~{a}o da famosa lei da radia\c{c}\~{a}%
o do corpo negro quando ``ap\'{o}s algumas semanas do mais extenuante
trabalho da minha vida, a escurid\~{a}o se desfez e uma inesperada vista come%
\c{c}ou a surgir''.\footnote{%
M. Planck, citado na Ref. \cite{klein62}, p. 468.}

Os trabalhos seminais sobre a quantiza\c{c}\~{a}o da energia foram
apresentados por Planck, perante a Academia Alem\~{a} de F\'{\i}sica, nas
sess\~{o}es de 19 de outubro e 14 de dezembro de 1900. Na primeira comunica%
\c{c}\~{a}o, uma nova f\'{o}rmula para a distribui\c{c}\~{a}o espectral da
radia\c{c}\~{a}o normal (corpo negro) \'{e} proposta. Na segunda, Planck
introduz a hip\'{o}tese de quantiza\c{c}\~{a}o da energia seguindo um m\'{e}%
todo n\~{a}o-ortodoxo inspirado nas id\'{e}ias da mec\^{a}nica estat\'{\i}%
stica de Ludwig Boltzmann (1844-1906). No in\'{\i}cio de 1901, aparece no 
\textit{Annalen der Physik} o trabalho completo, em que Planck apresenta uma
dedu\c{c}\~{a}o mais elaborada da sua distribui\c{c}\~{a}o espectral.%
\footnote{%
A vers\~{a}o para o portugu\^{e}s deste trabalho encontra-se neste n\'{u}%
mero da RBEF.}

\section{A Radia\c{c}\~{a}o do Corpo Negro}

Um corpo aquecido emite radia\c{c}\~{a}o eletromagn\'{e}tica em um largo
espectro cont\'{\i}nuo de freq\"{u}\^{e}ncias, principalmente na regi\~{a}o
do infravermelho (sensa\c{c}\~{a}o de calor), mas com intensidade vari\'{a}%
vel que atinge um m\'{a}ximo em um determinado comprimento de onda. \'{E}
bem conhecido, por exemplo, que um metal a 600$^{\text{o}}$C, (por exemplo,
em um forno el\'{e}trico) apresenta uma fraca colora\c{c}\~{a}o avermelhada
enquanto que o mesmo material (por exemplo, em uma sider\'{u}rgica)
apresenta uma cor azulada a temperaturas bem mais altas. O Sol, cuja
temperatura na superf\'{\i}cie \'{e} de cerca de 6.000$^{\text{o}}$C \'{e} o
exemplo mais familiar de emiss\~{a}o de radia\c{c}\~{a}o t\'{e}rmica, cujo
espectro abrange toda a regi\~{a}o v\'{\i}sivel incluindo a de comprimentos
de onda maiores (infravermelho) e menores (ultravioleta).

De uma maneira geral, mat\'{e}ria e radia\c{c}\~{a}o interagem e atingem o
equil\'{\i}brio termodin\^{a}mico atrav\'{e}s de trocas de energia. Sejam 
\emph{e} a pot\^{e}ncia emissiva, \textit{i.e}. a quantidade de energia
radiante emitida por unidade de \'{a}rea e por unidade de tempo, e \emph{a}
a absortividade ou absort\^{a}ncia, \textit{i.e}. a fra\c{c}\~{a}o de
energia incidente sobre a superf\'{\i}cie que \'{e} absorvida. W. Ritchie
(1833)\footnote{%
Citado na Ref. \cite{kangro76}, p. 7.} verificou o princ\'{\i}pio de
proporcionalidade entre emiss\~{a}o e absor\c{c}\~{a}o total, em sua famosa
experi\^{e}ncia com dois corpos radiantes $A$ e $B$, usando um term\^{o}%
metro diferencial. No equil\'{\i}brio t\'{e}rmico, o princ\'{\i}pio
estabelece que 
\begin{equation}
\frac{e_{A}}{a_{A}}=\frac{e_{B}}{a_{B}}.  \label{1}
\end{equation}
Suponha que um dos corpos apresente a especificidade de que $a_{N}=1$, ou
seja o corpo $N$ absorve toda a radia\c{c}\~{a}o que incide sobre ele. Da%
\'{\i}, denominamos $N$ de corpo negro. \'{E} evidente, da Eq. (\ref{1}),
que 
\begin{equation}
e_{N}=\frac{e_{A}}{a_{A}}\qquad \text{tal que\qquad }e_{N}>e_{A}  \label{2}
\end{equation}
para qualquer corpo $A$. Assim, o corpo negro possui uma pot\^{e}ncia
emissiva maior do que a de qualquer outro corpo. Evidentemente, um objeto
com estas caracter\'{\i}sticas \'{e} um corpo ideal que n\~{a}o pode ser
encontrado na pr\'{a}tica, mas pode ser constru\'{\i}do, numa boa aproxima%
\c{c}\~{a}o, atrav\'{e}s de uma caixa oca (um forno, por exemplo) com
paredes internas met\'{a}licas e uma pequena abertura que permite a passagem
de radia\c{c}\~{a}o, como ilustrado na Fig. \ref{Fig1}. A caixa deve ser
revestida de um excelente isolante t\'{e}rmico e espelhada externamente,
refletindo toda radia\c{c}\~{a}o eventualmente incidente, exceto na
abertura. A radia\c{c}\~{a}o, que entra na cavidade, tem uma probabilidade
muito pequena de escapar, permanecendo assim em seu interior e sendo
espalhada pelas paredes da cavidade at\'{e} atingir o equil\'{\i}brio t\'{e}%
rmico. Desta forma, toda a radia\c{c}\~{a}o incidente \'{e} absorvida pelo
corpo.

\begin{center}
\FRAME{fhFU}{2.9871in}{2.002in}{0pt}{\Qcb{Representa\c{c}\~{a}o esquem\'{a}%
tica de um corpo negro.}}{\Qlb{Fig1}}{fig-0101.tif}{\special{language
"Scientific Word";type "GRAPHIC";maintain-aspect-ratio TRUE;display
"ICON";valid_file "F";width 2.9871in;height 2.002in;depth 0pt;original-width
2.9438in;original-height 1.9631in;cropleft "0";croptop "1";cropright
"1";cropbottom "0";filename 'fig-0101.TIF';file-properties "XNPEU";}}
\end{center}

A radia\c{c}\~{a}o contida na cavidade pode ser decomposta em suas
componentes espectrais atrav\'{e}s de uma fun\c{c}\~{a}o distribui\c{c}\~{a}%
o $\rho (\nu ,T)$ tal que $\rho (\nu ,T)d\nu $ \'{e} a \textit{densidade de
energia (energia por unidade de volume) da radia\c{c}\~{a}o com freq\"{u}%
\^{e}ncia no intervalo compreendido entre }$\nu $\textit{\ e }$\nu +d\nu $
quando a cavidade est\'{a} a uma temperatura absoluta $T$. O espectro
emitido pela cavidade \'{e} especificado pelo fluxo de energia $R(\nu ,T)$
que, obviamente, deve ser proporcional a $\rho (\nu ,T)$, com a constante de
proporcionalidade advinda de fatores geom\'{e}tricos.\footnote{%
Para o c\'{a}lculo deste fator, ver Ref. \cite{reif65}, p. 387.} Na verdade,
temos 
\begin{equation}
R(\nu ,T)=\frac{c}{4}\rho (\nu ,T),  \label{3}
\end{equation}
em que $c$ \'{e} a velocidade da luz no v\'{a}cuo. \'{E} esta intensidade de
energia que \'{e} medida experimentalmente.

\subsection{Leis Universais do Espectro}

Em 1859, Gustav Robert Kirchhoff (1824-1887) apresentou perante a Academia
de Berlim o trabalho\ \textit{Sobre a rela\c{c}\~{a}o entre a emiss\~{a}o e
absor\c{c}\~{a}o de calor e luz}, em que provou que ``para raios de mesmo
comprimento de onda e a mesma temperatura, a raz\~{a}o entre a pot\^{e}ncia
emissiva e a absortividade s\~{a}o as mesmas para todos os corpos''.%
\footnote{%
Citado na Ref. \cite{jammer89}, p. 2.} O teorema foi demonstrado com ``base
em considera\c{c}\~{o}es te\'{o}ricas bastante simples'' e estabelece que,
para quaisquer corpos em equil\'{\i}brio t\'{e}rmico trocando radia\c{c}\~{a}%
o com comprimento de onda $\lambda $, a Eq. (\ref{1}) \'{e} satisfeita.%
\footnote{%
Deve ser mencionado, por raz\~{o}es hist\'{o}ricas, que Kirchhoff se
interessou pelo estudo dos processos de emiss\~{a}o e absor\c{c}\~{a}o a
partir de suas investiga\c{c}\~{o}es das raias do espectro solar.} Somente
em um segundo artigo, Kirchhoff introduziu a no\c{c}\~{a}o de ``um corpo 
\textit{perfeitamente negro}''\footnote{\'{E} curioso assinalar que o termo 
\textit{corpo negro} j\'{a} havia sido usado, n\~{a}o com este significado,
por Isaac Newton (1642-1727) no livro \textit{Optics} (1704). Ver Ref.\label%
{jammer89} \cite{jammer89}, p. 50.}, como fiz na se\c{c}\~{a}o I, e mostrou
que a pot\^{e}ncia emissiva de um corpo negro depende s\'{o} da temperatura
e da freq\"{u}\^{e}ncia da radia\c{c}\~{a}o, tal que 
\begin{equation}
e_{N}=F(\nu ,T),  \label{4}
\end{equation}
onde $F(\nu ,T)$ \'{e} uma fun\c{c}\~{a}o universal independente da forma,
tamanho e composi\c{c}\~{a}o qu\'{\i}mica do corpo.

Com base na termodin\^{a}mica e na teoria eletromagn\'{e}tica da radia\c{c}%
\~{a}o, \'{e} poss\'{\i}vel deduzir duas leis relativas \`{a} depend\^{e}%
ncia da radia\c{c}\~{a}o do corpo negro com a temperatura.

\noindent 1. A partir de resultados experimentais de J. Tyndall (1864) de
que a emiss\~{a}o total de um fio de platina a 1.200$^{\text{o}}$C \'{e} $%
11,7$ vezes maior que a correspondente emiss\~{a}o a 525$^{\text{o}}$C,
Josef Stefan (1835-1893) concluiu, em 1879,\ que a energia total \'{e}
proporcional \`{a} quarta pot\^{e}ncia da temperatura absoluta, pois
(1473/798)$^{4}\simeq 11,7.$ Este resultado fortuito\footnote{%
Na verdade, Tyndall mediu uma radiac\~{a}o que estava longe de ser a de um
corpo negro.} foi demonstrado rigorosamente por Boltzmann (1884) com base na
exist\^{e}ncia de uma press\~{a}o de radia\c{c}\~{a}o $-$ que mostrou ser $%
p=U/3$ usando a teoria eletromagn\'{e}tica de James Clerck Maxwell
(1831-1879) $-$, e considerando a radia\c{c}\~{a}o como uma m\'{a}quina t%
\'{e}rmica sujeita \`{a}s leis da termodin\^{a}mica para concluir que 
\begin{equation}
U=\sigma T^{4},  \label{5}
\end{equation}
onde $\sigma $ \'{e} a constante de Stefan-Boltzmann.\footnote{%
Veja a dedu\c{c}\~{a}o da lei $T^{4}$ na Ref. \cite{brush76}. Considere um g%
\'{a}s ideal de radia\c{c}\~{a}o com energia $E$, densidade de energia $U$ e
press\~{a}o $p=(1/3)U.$ \ Usando a rela\c{c}\~{a}o termodin\^{a}mica $\left(
\partial E/\partial V\right) _{T}=T\left( \partial p/\partial T\right)
_{V}-p $ e o fato que $\left( \partial E/\partial V\right) _{T}=U$, pois $U$ 
\'{e} uma fun\c{c}\~{a}o apenas da temperatura, segue, ap\'{o}s algumas
manipula\c{c}\~{o}es, que $dU/U=4(dT/T).$ Integrando obt\'{e}m-se o
resultado desejado.}

\noindent 2. A outra lei, chamada \emph{lei de deslocamento, }data de 1893,
e foi demonstrada por Wilhelm Wien (1864-1928). A lei estabelece que a
distribui\c{c}\~{a}o espectral da densidade de energia \'{e} dada pela equa%
\c{c}\~{a}o 
\begin{equation}
\rho (\nu ,T)=\nu ^{3}f(\frac{\nu }{T}),  \label{6}
\end{equation}
em que $f(\nu /T)$ \'{e} uma fun\c{c}\~{a}o apenas da raz\~{a}o entre a freq%
\"{u}\^{e}ncia e a temperatura. Esta rela\c{c}\~{a}o pode ser deduzida atrav%
\'{e}s do efeito Doppler que surge quando radia\c{c}\~{a}o incide sobre um
espelho hipot\'{e}tico m\'{o}vel.\footnote{%
Ver ap\^{e}ndice XXXIII da Ref.\cite{born}.} Observe que a lei de
Stefan-Boltzmann est\'{a} contida na lei de deslocamento de Wien, desde que 
\begin{equation*}
U=\int \rho (\nu ,T)d\nu =\int \nu ^{3}f(\frac{\nu }{T})d\nu =T^{4}\int
x^{3}f(x)dx=\sigma T^{4}.
\end{equation*}
A lei de Wien pode ser escrita, de uma outra forma, em termos do comprimento
de onda da radia\c{c}\~{a}o ao inv\'{e}s da freq\"{u}\^{e}ncia. Ora, $\rho
(\nu ,T)d\nu =\rho (\lambda ,T)d\lambda ,$ e como $\lambda \nu =c$, temos $%
\left| d\nu \right| /\nu =\left| d\lambda \right| /\lambda $. Portanto 
\begin{equation}
\rho (\lambda ,T)d\lambda =\frac{c^{4}}{\lambda ^{5}}f(\frac{c}{\lambda T}%
)d\lambda =\lambda ^{-5}\varphi (\lambda T)d\lambda .  \label{7}
\end{equation}
A origem do nome ``lei de deslocamento'' deve-se ao fato de que o
comprimento de onda, no qual a intensidade de radia\c{c}\~{a}o \'{e} m\'{a}%
xima, varia com a temperatura de acordo com a rela\c{c}\~{a}o 
\begin{equation}
\lambda _{\max }T=b\text{ \quad ou\quad }\lambda _{\max }\propto \frac{1}{T}.
\label{8}
\end{equation}
Esta rela\c{c}\~{a}o \'{e} f\'{a}cilmente deduzida da Eq. (\ref{7}),
simplesmente calculando o comprimento de onda para o qual $\rho (\lambda ,T)$
\'{e} m\'{a}ximo. Obviamente, o valor de $b$ depende da forma da fun\c{c}%
\~{a}o $f(c/\lambda T).$

A lei de Wien foi verificada experimentalmente in\'{u}meras vezes $-$ a
confirma\c{c}\~{a}o mais cuidadosa foi a de Friedrich Paschen (1865-1947) $-$%
, e constituiu um consider\'{a}vel avan\c{c}o pois permitia determinar a
distribui\c{c}\~{a}o espectral para qualquer temperatura, uma vez que se
conhecesse a distribui\c{c}\~{a}o em uma dada temperatura. Otto Lummer
(1860-1925) e Ernst Pringsheim (1859-1917), em Berlim, confirmaram a
validade da Eq. (\ref{8}), encontrando que $b=0,294$ cm.grau.\footnote{%
Planck vai usar este valor para determinar a constante $h$ e a constante de
Boltzmann $k_{B}$.}

\subsection{F\'{o}rmulas Emp\'{\i}ricas da Distribui\c{c}\~{a}o Espectral}

No entanto, nem os princ\'{\i}pios e rela\c{c}\~{o}es b\'{a}sicas da termodin%
\^{a}mica nem do eletromagnetismo, por si s\'{o}, permitem achar a forma
funcional de $f(\nu /T)$ ou $\varphi (\lambda T)$. Sua determina\c{c}\~{a}o
era um dos maiores problemas da f\'{\i}sica te\'{o}rica no final do s\'{e}%
culo XIX.

Uma das conjecturas propostas, em 1896 pelo pr\'{o}prio Wien\cite{wien97},
foi baseada em argumentos d\'{u}bios de que a distribui\c{c}\~{a}o de
energia deveria ser do tipo daquela proposta por Maxwell para a distribui%
\c{c}\~{a}o de velocidades de mol\'{e}culas de um g\'{a}s. Wien argumentou
que o n\'{u}mero de mol\'{e}culas \'{e} proporcional a $\exp (-mv^{2}/k_{B}T)
$ $-$ express\~{a}o que deveria ser v\'{a}lida tamb\'{e}m para mol\'{e}culas
no s\'{o}lido $-$ e ``uma vis\~{a}o atualmente aceita \'{e} que as cargas el%
\'{e}tricas das mol\'{e}culas podem excitar ondas eletromagn\'{e}ticas...[e]
como o comprimento de onda $\lambda $ da radia\c{c}\~{a}o emitida por uma
dada mol\'{e}cula \'{e} uma fun\c{c}\~{a}o de $v$, $v$ \'{e} tamb\'{e}m uma
fun\c{c}\~{a}o de $\lambda $''. Assim, usando a sua lei, dada pela Eq. (\ref%
{7}), prop\^{o}s que a distribui\c{c}\~{a}o espectral poderia ser dada por 
\begin{equation}
\varphi (\lambda T)=C\exp (-c/\lambda T)\text{\qquad ou\qquad }f(\frac{\nu }{%
T})=\alpha \exp (-\beta \frac{\nu }{T}),  \label{10}
\end{equation}%
onde $C,$ $c,$ $\alpha $ e $\beta $ s\~{a}o constantes. Merece ser destacado
o papel desempenhado pelo \textit{Physicalisch-Technische Reichsanstalt }e
os experimentos a\'{\i} realizados a partir de 1896 atrav\'{e}s das figuras
proeminentes de Otto Lummer, membro do laborat\'{o}rio desde 1889, e seus
colegas e colaboradores Ernst Pringsheim, Heinrich Rubens e Ferdinand
Kurlbaum. As medidas da distribui\c{c}\~{a}o espectral eram dif\'{\i}ceis de
serem obtidas com a precis\~{a}o necess\'{a}ria para decidir dentre v\'{a}%
rias f\'{o}rmulas emp\'{\i}ricas propostas \ Uma descri\c{c}\~{a}o detalhada
dos resultados experimentais pode ser encontrada na Ref. \cite{kangro76}. A
Fig. \ref{Fig2} mostra os resultados de Lummer e Pringsheim ao final de
1989. A f\'{o}rmula de Wien ajustava-se satisfatoriamente aos resultados
experimentais `preliminares'.

\FRAME{fhFU}{2.0081in}{3.0113in}{0pt}{\Qcb{Intensidade espectral como fun
\c{c}\~{a}o do comprimento de onda obtida por Lummer e Pringsheim em
novembro de 1899. Extra\'{\i}do da Ref. \protect\cite{kangro76}, p. 176.}}{%
\Qlb{Fig2}}{lummerbb.tif}{\special{language "Scientific Word";type
"GRAPHIC";maintain-aspect-ratio TRUE;display "USEDEF";valid_file "F";width
2.0081in;height 3.0113in;depth 0pt;original-width 3.6192in;original-height
6.4281in;cropleft "0";croptop "1.1376";cropright "1.3415";cropbottom
"0";filename 'lummerBB.tif';file-properties "XNPEU";}}

\section{A Teoria de Planck sobre a Radia\c{c}\~{a}o do Calor}

A partir do desafio de se encontrar uma f\'{o}rmula precisa e bem
fundamentada da distribui\c{c}\~{a}o espectral, nasceu a teoria qu\^{a}ntica
introduzida por Planck.

Em 1897, Planck, com quase 40 anos $-$ assim como Galileo (1546-1642), fez
suas descobertas com uma idade j\'{a} avan\c{c}ada se comparada com \`{a}%
quela de outros grandes cientistas na \'{e}poca de suas maiores contribui%
\c{c}\~{o}es $-$, come\c{c}ou a investigar o problema da radia\c{c}\~{a}o do
corpo negro. De acordo com o teorema de Kirchhoff, a radia\c{c}\~{a}o possu%
\'{\i}a um car\'{a}ter universal, de modo que Planck procurou uma abordagem
que se baseasse na eletrodin\^{a}mica e na irreversibilidade do processo que
conduzia a radia\c{c}\~{a}o ao equil\'{\i}brio t\'{e}rmico. A quest\~{a}o
principal seria determinar como a radia\c{c}\~{a}o e a mat\'{e}ria interagem
e atingem o equil\'{\i}brio. Como podemos explicar que um sistema
conservativo formado de radia\c{c}\~{a}o eletromagn\'{e}tica e uma cole\c{c}%
\~{a}o de osciladores harm\^{o}nicos $-$ que Planck chamou ressonadores $-$
chega ao equil\'{\i}brio sem invocar outras hip\'{o}teses al\'{e}m das leis
da teoria eletromagn\'{e}tica e da termodin\^{a}mica? A escolha dos
osciladores harm\^{o}nicos, como um modelo simples para a mat\'{e}ria, pode
ser justificada pelo teorema de Kirchhoff que assegura a independ\^{e}ncia
da distribui\c{c}\~{a}o da radia\c{c}\~{a}o do corpo negro em rela\c{c}\~{a}%
o \`{a} composi\c{c}\~{a}o da mat\'{e}ria.

Em 1899, Planck provou um teorema bastante importante que estabelece uma rela%
\c{c}\~{a}o entre a densidade de energia $\rho (\nu ,T)$ e a energia m\'{e}%
dia $\overline{u}(\nu ,T)$ do conjunto de osciladores harm\^{o}nicos que
representava os \'{a}tomos na superf\'{\i}cie interna da cavidade no corpo
negro em equil\'{\i}brio termodin\^{a}mico. O resultado pode ser escrito como%
\footnote{%
Deriva\c{c}\~{o}es simplificadas encontram-se no Ap\^{e}ndice A da Ref. \cite%
{jammer89}, p. 407 e Ref. \cite{pais79}, p. 870.} 
\begin{equation}
\rho (\nu ,T)=\frac{8\pi \nu ^{2}}{c^{3}}\overline{u}(\nu ,T).  \label{11}
\end{equation}
Como a radia\c{c}\~{a}o e os osciladores est\~{a}o em equil\'{\i}brio, a freq%
\"{u}\^{e}ncia $\nu $ tem duplo significado: representa a freq\"{u}\^{e}ncia
da radia\c{c}\~{a}o incidente assim como uma poss\'{\i}vel freq\"{u}\^{e}%
ncia dos modos de oscila\c{c}\~{a}o dos \'{a}tomos na parede da cavidade.

A Eq. (\ref{11}) pode ser obtida, em linhas gerais, pela determina\c{c}\~{a}%
o da energia irradiada por segundo por um carga acelerada. Este \'{e} um c%
\'{a}lculo essencial na teoria eletromagn\'{e}tica e pode ser encontrado em
in\'{u}meros livros-textos. O resultado \'{e} importante para a compreens%
\~{a}o das propriedades da radia\c{c}\~{a}o eletromagn\'{e}tica, n\~{a}o
apenas no corpo negro, mas tamb\'{e}m daquela emitida por \'{a}tomos, esta%
\c{c}\~{o}es de r\'{a}dio, estrelas e at\'{e} na origem da cor azul dos c%
\'{e}us.\cite{crawford} A famosa f\'{o}rmula para a pot\^{e}ncia irradiada
por um dipolo oscilante \'{e} dada por $P(t)=\frac{2}{3}\frac{q^{2}}{c^{3}}%
\mathbf{a}^{2},$ em que $\mathbf{a}$ \'{e} a acelera\c{c}\~{a}o da carga $q$%
. No caso de oscilador harm\^{o}nico (dipolo oscilante), $a=-(2\pi \nu
)^{2}x,$ e tomando-se a m\'{e}dia temporal, transforma-se em $P=\frac{2}{3}%
\frac{q^{2}}{c^{3}}(2\pi \nu )^{4}\left\langle x^{2}\right\rangle $ em que o
valor m\'{e}dio \'{e} tomado em intervalos de tempo longos comparados com o
per\'{\i}odo de oscila\c{c}\~{a}o, mas suficientemente pequenos para
desprezarmos a radia\c{c}\~{a}o nestes intervalos. Como a energia total m%
\'{e}dia dos osciladores $\overline{u}=m(2\pi \nu )^{2}\left\langle
x^{2}\right\rangle $, ent\~{a}o $P=\frac{2}{3}\frac{q^{2}}{mc^{3}}(2\pi \nu
)^{2}\overline{u}.$ Por outro lado, o trabalho fornecido por segundo ao
oscilador por um campo de radia\c{c}\~{a}o com densidade de energia $\rho
(\nu )$ \'{e} dado por $P=\frac{\pi q^{2}}{3m}\rho (\nu ),$ que resulta da
solu\c{c}\~{a}o da equa\c{c}\~{a}o do movimento do oscilador harm\^{o}nico
na presen\c{c}a de um campo el\'{e}trico com freq\"{u}\^{e}ncia $\nu .$%
\footnote{%
A dedu\c{c}\~{a}o pode ser vista no Ap\^{e}ndice XXXIV da Ref. \cite{born}.}
Igualando as duas express\~{o}es, temos o resultado do teorema de Planck,
Eq. (\ref{11}).

\'{E} evidente que Planck precisava calcular a energia m\'{e}dia de um
oscilador harm\^{o}nico a uma temperatura $T$ para determinar, atrav\'{e}s
da Eq. (\ref{11}), a distribui\c{c}\~{a}o espectral. Poderia usar o
resultado j\'{a} conhecido do teorema da equiparti\c{c}\~{a}o da energia $-$
que n\~{a}o levaria \`{a} resposta correta como discutiremos a seguir. No
entanto, Planck preferiu usar uma abordagem ``termodin\^{a}mica'', talvez
devido ao seu continuado interesse nesta linha de pesquisa desde o seu
doutorado, cuja tese consistiu numa rean\'{a}lise do trabalho de Rudolf
Clausius (1822-1888) sobre a segunda lei da termodin\^{a}mica em termos da no%
\c{c}\~{a}o de entropia.

Usando a f\'{o}rmula de Wien, dada pela Eqs. (\ref{6}) e (\ref{10}), e a Eq.
(\ref{11}), temos $\rho (\nu ,T)=(\alpha c^{3}/8\pi )\nu \exp (-\beta \nu
/T).$ Deste resultado, temos que $T^{-1}=-\frac{1}{\beta \nu }\ln (\frac{%
8\pi }{\alpha c^{3}}\frac{u}{\nu }).$ Mas, como $T^{-1}=\partial S/\partial
u $ (a volume constante), a equa\c{c}\~{a}o pode ser integrada e a entropia
do oscilador $S$ pode ser escrita em termos de sua energia $u$ [antes
denotada por $\overline{u}(\nu ,T)$] como 
\begin{equation}
S=-\frac{u}{\beta \nu }\ln \frac{u}{Ae\nu },  \label{15}
\end{equation}
em que $A=\alpha c^{3}/8\pi $ e $e$ \'{e} a base do logaritmo natural. Com a
entropia assim definida, Planck determinou a entropia da radia\c{c}\~{a}o em
equil\'{\i}bro com o conjunto de osciladores e mostrou que esta satisfazia a
segunda lei da termodin\^{a}mica. Mais ainda, Planck ficou impressionado com
a simplicidade da Eq. (\ref{15}) que implicava que $\partial ^{2}S/\partial
u^{2}\propto -u^{-1}$.

Planck mostrou ainda que qualquer outra f\'{o}rmula proposta para $\rho (\nu
,T)$ deveria ser tal que $\partial ^{2}S/\partial u^{2}$ fosse uma fun\c{c}%
\~{a}o negativa \ da energia $u$ de modo a satisfazer a segunda lei da
termodin\^{a}mica.

\section{A F\'{o}rmula Emp\'{\i}rica de Planck}

No in\'{\i}cio de 1900, as duas equipes do \textit{Physicalisch-Technische
Reichsanstalt }em Berlim, formadas por Lummer e Pringsheim e Rubens e
Kurlbaum, independentemente conseguiram medir a radia\c{c}\~{a}o numa regi%
\~{a}o ainda inexplorada de grande comprimentos de onda. O primeiro time
varreu a regi\~{a}o de $\lambda =12-18$ $\mu $m, e $T=300-1.650$ K,
concluindo que a f\'{o}rmula de Wien n\~{a}o era v\'{a}lida para estes
valores de comprimentos de onda mais longos. Em outubro de 1900, o trabalho
experimental muito cuidadoso de Rubens e Kurlbaum na regi\~{a}o do
infravermelho mais long\'{\i}nguo $\lambda =30-60$ $\mu $m, e $T=200-1.500^{%
\text{o}}$C mostrava, sem nenhuma d\'{u}vida, que, para comprimentos de onda
longos dentro de uma grande faixa de temperatura, a f\'{o}rmula de Wien era
inadequada. Kurlbaum apresentou estes resultados na mesma sess\~{a}o da
Academia Alem\~{a} de F\'{\i}sica, de 19 de outubro, em que Planck
apresentou a sua f\'{o}rmula.

A Fig. \ref{Fig3} mostra os pontos experimentais da intensidade de radia\c{c}%
\~{a}o do corpo negro como fun\c{c}\~{a}o da temperatura para $\lambda =51,2$
$\mu $m comparados com as curvas relativas \`{a}s f\'{o}rmulas de Wien, Eq. (%
\ref{10}), de Thiesen (que n\~{a}o discutirei neste artigo), de Lord
Rayleigh, Eq. (\ref{22}), e a de Planck, Eq. (\ref{18}). Em 7 de outubro,
Rubens visitou Planck e lhe informou que, para longos comprimentos de onda
ou baixas freq\"{u}\^{e}ncias, $\rho (\nu ,T)\propto T$. Planck descobriu a
sua f\'{o}rmula da radia\c{c}\~{a}o neste dia, e informou a Rubens seu
resultado, atrav\'{e}s de um cart\~{a}o postal na noite de mesmo dia.%
\footnote{%
Ver Ref. \cite{kangro76}, p. 205.}

\FRAME{fhFU}{3.0874in}{2.6498in}{0pt}{\Qcb{Curvas da energia da radia\c{c}\~%
{a}o versus temperatura, medida atrav\'{e}s dos raios residuais
(``reststrahlen'') usando pedras de sal ($\protect\lambda =51,2$ $\protect%
\mu $m), e comparados (``berechnet nach'' significa ``calculado ap\'{o}s'')
com as f\'{o}rmulas de Wien, Lord Rayleigh, Thiesen e Planck. Ref. %
\protect\cite{kangro76}, p. 204.}}{\Qlb{Fig3}}{rubens-sal.tif}{\special%
{language "Scientific Word";type "GRAPHIC";maintain-aspect-ratio
TRUE;display "USEDEF";valid_file "F";width 3.0874in;height 2.6498in;depth
0pt;original-width 5.6896in;original-height 4.8767in;cropleft "0";croptop
"1";cropright "1";cropbottom "0";filename 'rubens-sal.tif';file-properties
"XNPEU";}}

Em sua comunica\c{c}\~{a}o em 19 de outubro,\footnote{%
A vers\~{a}o em portugu\^{e}s \'{e} o artigo seguinte deste n\'{u}mero da
RBEF.} Planck apresentou a sua f\'{o}rmula para a distribui\c{c}\~{a}o
espectral da radia\c{c}\~{a}o do corpo negro, obtida pela interpola\c{c}\~{a}%
o entre os resultados previstos para $\rho (\nu ,T)$ nos limites extremos da
freq\"{u}\^{e}ncia. Para altas freq\"{u}\^{e}ncias, $\partial ^{2}S/\partial
u^{2}\propto -u^{-1}$ e para baixas freq\"{u}\^{e}ncias $\partial
^{2}S/\partial u^{2}\propto -u^{-2}$, como pode ser facilmente visto, j\'{a}
que se $\rho (\nu ,T)\propto T$, ent\~{a}o tamb\'{e}m $u\propto T$ . Usando $%
T^{-1}=\partial S/\partial u$, chega-se \`{a} depend\^{e}ncia desejada.
Planck ent\~{a}o prop\^{o}s uma express\~{a}o ``quase t\~{a}o simples quanto
a express\~{a}o de Wien, e que mereceria ser investigada uma vez que a
express\~{a}o de Wien n\~{a}o \'{e} suficiente para cobrir todas as observa%
\c{c}\~{o}es'' dada por 
\begin{equation}
\frac{\partial ^{2}S}{\partial u^{2}}=-\frac{1}{u(\alpha +u)},  \label{16}
\end{equation}
em que ``uso a derivada segunda de $S$ em rela\c{c}\~{a}o a $u$ porque esta
quantidade tem um significado f\'{\i}sico simples. Esta \'{e}, de longe, a
mais simples de todas as express\~{o}es que leva $S$ a ser uma fun\c{c}\~{a}%
o logar\'{\i}tmica de $U$''. Integrando a Eq. (\ref{16}), temos $\partial
S/\partial u=(1/\alpha )\ln [(\alpha +u)/u)]+c,$ onde $c$ \'{e} uma
constante arbitr\'{a}ria. Usando $\partial S/\partial u=T^{-1}$, $(1/\alpha
)\ln [(\alpha +u)/u)]+c=T^{-1}$, e encontramos $c=0$, porque no limite de
altas temperaturas ambos os lados da equa\c{c}\~{a}o devem se anular. Assim,
a energia do oscilador \'{e} dada por 
\begin{equation}
u=\frac{\alpha }{e^{\alpha /T}-1},  \label{17}
\end{equation}
e usando a Eq. (\ref{11}), a distribui\c{c}\~{a}o de energia vem a ser dada
pela express\~{a}o 
\begin{equation*}
\rho (\nu ,T)=\frac{8\pi }{c^{3}}\frac{\alpha \nu ^{2}}{e^{\alpha /T}-1}.
\end{equation*}
A lei de deslocamento de Wien, Eq. (\ref{6}) torna claro que $\alpha $ deve
ser uma fun\c{c}\~{a}o linear de $\nu $. Uma express\~{a}o geral, em termos
de duas constantes gen\'{e}ricas $A$ e $B$, pode ser escrita como 
\begin{equation}
\rho (\nu ,T)=\frac{8\pi }{c^{3}}\frac{A\nu ^{3}}{e^{B\nu /T}-1}.  \label{18}
\end{equation}
E Planck conlui: ``Assim, permiti-me chamar a sua aten\c{c}\~{a}o para esta
nova f\'{o}rmula, que considero ser, exceto a express\~{a}o de Wien, a mais
simples poss\'{\i}vel do ponto de vista da teoria eletromagn\'{e}tica da
radia\c{c}\~{a}o.''

\section{A Lei ``Cl\'{a}ssica'' da Radia\c{c}\~{a}o T\'{e}rmica}

Farei aqui uma pequena digress\~{a}o, por raz\~{o}es de completitude hist%
\'{o}rica, para assinalar a contribui\c{c}\~{a}o de Lord Rayleigh (John
William Strutt, 1842-1919) \`{a} investiga\c{c}\~{a}o da radia\c{c}\~{a}o do
corpo negro que se tornou marcante como sendo o resultado cl\'{a}ssico da
distribui\c{c}\~{a}o espectral, baseado na mec\^{a}nica estat\'{\i}stica cl%
\'{a}ssica de Maxwell-Boltzmann.

Em uma curta nota, publicada em junho de 1900, Rayleigh\cite{rayleigh00}
aplicou a ``doutrina de Maxwell-Boltzmann da parti\c{c}\~{a}o da energia'', 
\textit{i.e}., o teorema da equiparti\c{c}\~{a}o da energia \`{a}s oscila%
\c{c}\~{o}es eletromagn\'{e}ticas da radia\c{c}\~{a}o na cavidade e
encontrou uma f\'{o}rmula radicalmente contr\'{a}ria \`{a} f\'{o}rmula de
Wien. Seu m\'{e}todo consistia em calcular o n\'{u}mero de ondas estacion%
\'{a}rias, ou seja a distribui\c{c}\~{a}o de modos eletromagn\'{e}ticos
permitidos com freq\"{u}\^{e}ncia no intervalo entre $\nu $ e $\nu +d\nu $, $%
\mathcal{N}(\nu )d\nu $, dentro da cavidade. \'{E} bem conhecido que%
\footnote{%
Veja por exemplo, Ref. \cite{eisberg79}, p. 45.} 
\begin{equation}
\mathcal{N}(\nu )d\nu =\frac{8\pi V}{c^{3}}\nu ^{2}d\nu .  \label{19}
\end{equation}
Para encontrar a densidade de energia, devemos saber a energia m\'{e}dia de
cada oscilador. O\emph{\ }teorema da equiparti\c{c}\~{a}o da energia
estabelece que cada termo da energia proporcional ao quadrado da coordenada,
momentum (ou amplitude da onda) contribui sempre com a mesma quantidade para
a energia m\'{e}dia, exatamente ($1/2)k_{B}T$. Lembre-se que no caso do
oscilador $\varepsilon =p^{2}/2m+(1/2)m\omega ^{2}x^{2}$, e no caso da radia%
\c{c}\~{a}o eletromagn\'{e}tica $\varepsilon \propto (E_{0}^{2}+B_{0}^{2})$.
Assim, teremos 
\begin{equation}
\overline{u}(T)=k_{B}T.  \label{20}
\end{equation}
A lei obtida por Rayleigh para a radia\c{c}\~{a}o do corpo negro pode ser
expressa atrav\'{e}s do produto do n\'{u}mero de ondas eletromagn\'{e}ticas
dentro da cavidade pela energia de cada uma delas. O resultado \'{e} 
\begin{equation}
\rho (\nu ,T)d\nu =\frac{\mathcal{N}(\nu )d\nu }{V}\times \overline{u}(T)=%
\frac{8\pi }{c^{3}}(k_{B}T)\nu ^{2}d\nu .  \label{21}
\end{equation}
A lei de radia\c{c}\~{a}o de Rayleigh \'{e} conhecida como \textit{lei de
Rayleigh-Jeans}, ap\'{o}s a contribui\c{c}\~{a}o de James Jeans (1877-1946),
em maio de 1905, ao introduzir o fator $1/8$ no c\'{a}lculo de $\mathcal{N}%
(\nu )d\nu ,$ que fora esquecido por Rayleigh em um trabalho de 1905.%
\footnote{%
Neste ano, Rayleigh e Jeans trocaram v\'{a}rias cartas na Nature. Ver
cronologia de eventos na Ref. \cite{pais79}, p. 872.} Como o seu resultado
era consideravelmente diferente da aclamada f\'{o}rmula de Wien, Rayleigh
introduziu um fator exponencial, tal que a express\~{a}o completa modificada 
\'{e} 
\begin{equation}
\rho (\nu ,T)d\nu =c_{1}T\nu ^{2}\exp (-c_{2}\frac{\nu }{T}).  \label{22}
\end{equation}
Rayleigh conclui sua nota\footnote{%
Na verdade, os fatores num\'{e}ricos s\'{o} foram calculados no artigo de
1905, sem o termo exponencial.} com o seguinte coment\'{a}rio: ``Se a Eq. (%
\ref{22}) representa as observa\c{c}\~{o}es, eu n\~{a}o estou em posi\c{c}%
\~{a}o de afirmar. Espera-se que uma resposta a esta quest\~{a}o pode ser
encontrada brevemente pelas m\~{a}os de destacados experimentais que t\^{e}m
se ocupado deste assunto.''

Digno de registro \'{e} o fato que Planck poderia ter usado o teorema da
equiparti\c{c}\~{a}o da energia, $\overline{u}(T)=k_{B}T$, na sua Eq. (\ref%
{11}), e obtido o mesmo resultado. Isto foi feito por Albert Einstein
(1879-1955) em seu trabalho de 1905, em que introduziu o conceito de quantum
de luz.\cite{einstein05}

Planck n\~{a}o se referiu a este resultado de Rayleigh em seus trabalhos,
mas obviamente o conhecia, porque fora publicado na prestigiosa \textit{%
Philosophical Magazine} e Rubens, na visita domiciliar de 7 de outubro, lhe
comunicara que, para baixas freq\"{u}\^{e}ncias, o resultado observado
correspondia \`{a} f\'{o}rmula de Lord Rayleigh, dada pela Eq. (\ref{22}).

Embora a lei de Rayleigh-Jeans, Eq. (\ref{21}), satisfa\c{c}a a lei de
deslocamento de Wien, dada pela Eq. (\ref{6}) com $f(\nu /T)=(\nu /T)^{-1}$,
a f\'{o}rmula falha no limite de grandes freq\"{u}\^{e}ncias e conduz a uma
diverg\^{e}ncia na densidade de energia total, como tamb\'{e}m apontado por
Einstein,\cite{einstein05} 
\begin{equation*}
U=\int \rho (\nu ,T)d\nu \propto \int\limits_{0}^{\infty }\nu ^{2}dv=\infty .
\end{equation*}
Este resultado ficou conhecido posteriormente como a ``cat\'{a}strofe do
ultravioleta'', gra\c{c}as a Paul Ehrenfest (1880-1933).\footnote{%
O termo apareceu pela primeira vez no seu quarto cap\'{\i}tulo de seu artigo
publicado no Ann. Phys. \textbf{36}, 91 (1911), reimpresso na Ref. \cite%
{ehrenfest14}.}

\section{A Introdu\c{c}\~{a}o dos Quanta}

Na reuni\~{a}o de 14 de dezembro de 1900, Planck comunicou aos membros da
Sociedade Alem\~{a} de F\'{\i}sica a dedu\c{c}\~{a}o te\'{o}rica de sua f%
\'{o}rmula, proposta em 19 de outubro, e no que veio a chamar de um ``ato de
desepero'' teve que introduzir a hip\'{o}tese da descontinuidade da energia
dos osciladores. A nota \'{e} curta e de dif\'{\i}cil compreens\~{a}o. \cite%
{planck00-14} Tr\^{e}s semanas depois, Planck enviou um trabalho completo
para o \textit{Annalen der Physik}, onde apresentava uma dedu\c{c}\~{a}o
mais aprimorada de sua express\~{a}o para a distribui\c{c}\~{a}o espectral.

Nos dois meses que separaram as duas reuni\~{o}es, Planck mudou radicalmente
a sua linha de pensamento, exposta nos trabalhos anteriores, ao adotar as id%
\'{e}ias de Boltzmann acerca da rela\c{c}\~{a}o entre entropia e
probabilidade. Mais ainda, teve que inventar um dos conceitos mais b\'{a}%
sicos da teoria f\'{\i}sica. E o fez usando um m\'{e}todo n\~{a}o-ortodoxo
claramente diferente daquele empregado por Boltzmann.

Planck havia mostrado nos trabalhos anteriores que um ponto chave, para uma
teoria do espectro de radia\c{c}\~{a}o t\'{e}rmica, era a determina\c{c}\~{a}%
o te\'{o}rica da entropia em fun\c{c}\~{a}o da energia de um oscilador harm%
\^{o}nico com freq\"{u}\^{e}ncia $\nu ,$ como na Eq. (\ref{15}) no caso de
uso da f\'{o}rmula de Wien.\footnote{%
Esta tese \'{e} defendida por Klein. Conferir a Ref. \cite{klein62}, p. 469.}
Se sua espress\~{a}o para $\rho (\nu ,T$), Eq. (\ref{18}), estivesse
correta, ent\~{a}o, seguindo os mesmos passos na obten\c{c}\~{a}o da Eq. (%
\ref{15}), poder-se-ia obter a entropia do oscilador. Como, em sua f\'{o}%
rmula, $u=A^{\prime }\nu \lbrack \exp (B\nu /T)-1]^{-1}$, invertendo esta
equa\c{c}\~{a}o para $T^{-1}=\partial S/\partial U$, e ent\~{a}o integrando,
chega-se ao resultado 
\begin{equation}
S=\frac{A^{\prime }}{B}\left[ (1+\frac{u}{A^{\prime }\nu })\ln (1+\frac{u}{%
A^{\prime }\nu })-\frac{u}{A^{\prime }\nu }\ln \frac{u}{A^{\prime }\nu }%
\right] ,  \label{23}
\end{equation}
em que $A$ e $B$ s\~{a}o as constantes que aparecem na Eq. (\ref{18}) e $%
A^{\prime }=Ac^{3}/8\pi ^{2}.$ Para deduzir formalmente a Eq. (\ref{23}),
Planck tinha que procurar outro m\'{e}todo e o encontrou no trabalho de
Boltzmann.

Segundo Boltzmann, a entropia de um sistema em um dado estado \'{e}
proporcional \`{a} probabilidade daquele estado que, em nota\c{c}\~{a}o
moderna, pode ser escrita como\footnote{%
Na verdade, apesar desta f\'{o}rmula constar da l\'{a}pide de seu t\'{u}%
mulo, Boltzmann nunca a escreveu nesta forma. Quem o fez, foi Planck em seu
artigo de 1901 no \textit{Annalen der Physik}. Mais ainda, quem realmente
introduziu a constante $k_{B}$ foi Planck. \ ''Em v\'{a}rias ocasi\~{o}es,
nos \'{u}ltimos anos de vida, Planck comentou que, embora a constante fosse
compreens\'{\i}velmente conhecida como a constante de Boltzmann, este nunca
lhe atribu\'{\i}ra algum significado f\'{\i}sico nem nunca procurou estimar
o seu valor num\'{e}rico'' [\cite{klein62}, p. 471]. Planck ainda deu grande 
\^{e}nfase ao seu car\'{a}ter universal, em sua comunica\c{c}\~{a}o de 14 de
dezembro de 1900.} 
\begin{equation}
S=k_{B}\ln W,  \label{24}
\end{equation}
em que $W$ \'{e} o n\'{u}mero de ``complexos'' - como chamou Planck -, 
\textit{i.e}. o n\'{u}mero de arranjos microsc\'{o}picos compat\'{\i}veis
com o dado estado macrosc\'{o}pico, e $k_{B}$ \'{e} a constante de
Boltzmann. Como determinar $W?$ Planck argumenta que:

\begin{quotation}
{\small Ent\~{a}o, \`{a} energia total } 
\begin{equation}
u_{N}=Nu  \label{25}
\end{equation}
{\small de um tal sistema, formado por }$N${\small \ ressonadores,
corresponde uma certa entropia total } 
\begin{equation}
S_{N}=NS  \label{26}
\end{equation}
{\small do mesmo sistema, em que }$S${\small \ representa a entropia m\'{e}%
dia de um ressonador particular. Esta entropia }$S_{N}${\small \ depende da
desordem com a qual a energia total }$u_{N}${\small \ se reparte entre os
diferentes ressonadores individuais.}
\end{quotation}

{\small Pr\'{o}ximo passo:}

\begin{quotation}
{\small Importa agora encontrar a probabilidade }$W${\small , de modo que os 
}$N${\small \ ressonadores possuam em conjunto a energia total }$u_{N}$%
{\small . Para isto, ser\'{a} necess\'{a}rio que }$u_{N}${\small \ n\~{a}o
seja uma quantidade cont\'{\i}nua, infinitamente divis\'{\i}vel, mas antes
uma grandeza discreta, composta de um n\'{u}mero inteiro de partes finitas
iguais. Denominemos }$\varepsilon ${\small \ a tal parte elementar de
energia; teremos, portanto: } 
\begin{equation}
U_{N}=P\varepsilon ,  \label{27}
\end{equation}
{\small onde }$P${\small \ representa um n\'{u}mero inteiro, em geral
grande. Deixaremos, no momento, indeterminado o valor de }$\varepsilon $%
{\small .}
\end{quotation}

Planck diz que ``a an\'{a}lise combinat\'{o}ria'' mostra que o n\'{u}mero de
reparti\c{c}\~{o}es poss\'{\i}veis \'{e} 
\begin{equation}
W=\frac{(N+P-1)!}{(N-1)!P!}.  \label{28}
\end{equation}

Uma dedu\c{c}\~{a}o simplificada desta f\'{o}rmula foi dada por Ehrenfest e
Onnes (1914).\cite{ehrenfest14} A Eq. (\ref{28}) expressa o n\'{u}mero de
maneiras que $N$ ressonadores $R_{1},$ $R_{2},...$ $R_{N},$ podem ser
distribu\'{\i}dos pelos v\'{a}rios graus de energia determinados pela s\'{e}%
rie de m\'{u}ltiplos $0,$ $\varepsilon ,$ $2\varepsilon ...$ Considere um
exemplo especial para introduzirmos um s\'{\i}mbolo para a distribui\c{c}%
\~{a}o: $N=4$ e $P=7.$ Uma das poss\'{\i}veis distribui\c{c}\~{o}es \'{e}: $%
R_{1}$ tem energia $4\varepsilon ,$ $R_{2}$ tem energia $2\varepsilon ,$ $%
R_{3}$ tem $0\varepsilon $ e $R_{4}$ tem $4\varepsilon .$ O s\'{\i}mbolo
para esta distribui\c{c}\~{a}o, lido da esquerda para a direita, indica a
energia de $R_{1},$ $R_{2},R_{3},R_{4}$, na distribui\c{c}\~{a}o escolhida
que tem $u=7\varepsilon $, pode ser escrito como $\Vert \ \varepsilon \
\varepsilon \ \varepsilon \ \varepsilon \ \blacksquare \ \varepsilon \
\varepsilon \ \blacksquare \ \blacksquare \ \varepsilon \ \Vert $ Para
valores gerais de $N $ e $P$, o s\'{\i}mbolo ter\'{a} $P$ vezes o sinal $%
\varepsilon $ e ($N-1)$ vezes o sinal $\blacksquare $. A quest\~{a}o \'{e}
saber quantos s\'{\i}mbolos \textit{diferentes} podem ser formados na
maneira indicada do n\'{u}mero dado de $\varepsilon $ e $\blacksquare $. 
\'{E} evidente que os $(N-1+P) $ elementos $\varepsilon $ e $\blacksquare $
podem ser arranjados de $(N+P-1)!$ maneiras diferentes entre os terminais $%
\Vert \ \Vert $. Mas, \'{e} f\'{a}cil de ver, que cada vez, $(N-1)!P!$ das
combina\c{c}\~{o}es poss\'{\i}veis dar\~{a}o o mesmo s\'{\i}mbolo para a
distribui\c{c}\~{a}o (combina\c{c}\~{o}es que s\~{a}o formadas permutando os 
$P$ elementos $\varepsilon $ ou os $(N-1)$ elementos $\blacksquare $).
Assim, o resultado final \'{e} a divis\~{a}o dos dois termos.

O c\'{a}lculo da entropia agora \'{e} direto. Usando a f\'{o}rmula de
Stirling, $W=(N+P)^{N+P}/N^{N}P^{P},$\footnote{%
Rosenfeld \cite{rosenfeld36} sugere a seguinte linha de racioc\'{\i}nio de
Planck: Se a entropia dos $N$ osciladores era do tipo $k_{B}\log W$, ent\~{a}%
o para se obter a Eq. (\ref{23}), $W$ deveria ser uma express\~{a}o do tipo (%
$N+P)^{N+P}/N^{N}P^{P}.$} \ e a Eq. (\ref{24}), a entropia do ressonador em
fun\c{c}\~{a}o da sua energia \'{e} escrita como 
\begin{equation}
S=k_{B}\left[ (1+\frac{u}{\varepsilon })\ln (1+\frac{u}{\varepsilon })-\frac{%
u}{\varepsilon }\ln \frac{u}{\varepsilon }\right] .  \label{29}
\end{equation}
Observe que esta express\~{a}o \'{e} exatamente igual \`{a} Eq. (\ref{23}).
At\'{e} aqui, o tamanho dos elementos $\varepsilon $ \'{e} completamente
arbitr\'{a}rio. Contudo, $S$ deve depender de $\nu ,$ al\'{e}m de $u,$ e
como $k_{B}$ \'{e} uma constante universal, a depend\^{e}ncia com a freq\"{u}%
\^{e}ncia deve aparecer em $\varepsilon $.\footnote{%
Planck mostrou que, uma outra forma de se escrever a lei de Wien, \'{e} $%
S=f(u/\nu ).$ [Ver Eq. (10) de seu artigo de 1901].} Usando $\partial
S/\partial U=1/T,$ Planck encontra a energia m\'{e}dia dos osciladores como 
\begin{equation*}
u=\frac{\varepsilon }{e^{\varepsilon /k_{B}T}-1}.
\end{equation*}
De modo a satisfazer a lei de Wien, o elemento de energia $\varepsilon $
deve ser proporcional \`{a} freq\"{u}\^{e}ncia do oscilador 
\begin{equation}
\varepsilon =h\nu ,  \label{30}
\end{equation}
em que $h$ \'{e} a segunda constante universal da teoria. Usando a Eq. (\ref%
{11}), chega-se ao mesmo resultado obtido no trabalho anterior de Planck: 
\begin{equation}
\rho (\nu ,T)=\frac{8\pi }{c^{3}}\frac{h\nu ^{3}}{e^{h\nu /k_{B}T}-1}.
\label{31}
\end{equation}

A parte final de seu trabalho \'{e} destinada a obter os valores num\'{e}%
ricos das constantes $h$ e $k_{B}$ a partir dos resultados experimentais: a
constante de Stefan $\sigma $ \'{e} uma combina\c{c}\~{a}o de $h$ e $k_{B}$,
e a raz\~{a}o de $\lambda _{\max }T$ fornece uma segunda equa\c{c}\~{a}o
para $h$ e $k_{B}$. Encontrou $h=6,55\times 10^{-27}$ erg.s e $%
k_{B}=1,346\times 10^{-16}$ erg/K.

Na comunica\c{c}\~{a}o de 14 de dezembro de 1900, Planck comenta que podiam
ser deduzidas de sua teoria ``outras rela\c{c}\~{o}es...que parecem, para
mim, ser de consider\'{a}vel import\^{a}ncia para outros campos da f\'{\i}%
sica e tamb\'{e}m da qu\'{\i}mica''. Possivelmente, uma refer\^{e}ncia \`{a}
determina\c{c}\~{a}o da constante de Boltzmann, porque da entropia de um g%
\'{a}s ideal mostra-se que $k_{B}=R/N_{0}$, onde $R$ \'{e} a cosntante dos
gases e $N_{0}$ o n\'{u}mero de Avogadro. Como $R$ era bem conhecida, Planck
conseguiu encontrar o melhor valor para $N_{0}$, que na \'{e}poca, s\'{o}
era estimado indiretamente a partir de modelos ultra simplificados da teoria
cin\'{e}tica.

Planck determinou ainda, a partir de sua teoria, a carga do rec\'{e}%
m-descoberto el\'{e}tron de acordo com a equa\c{c}\~{a}o $e=F(k_{B}/R),$
onde $F$ \'{e} a constante de Faraday, a carga de um \'{a}tomo-grama de 
\'{\i}ons monovalentes. Ele achou o valor $e=4,69\times 10^{-10}$ esu.
Planck ressaltou a import\^{a}ncia da determina\c{c}\~{a}o destas constantes
b\'{a}sicas que a sua teoria tornou poss\'{\i}vel.

\section{Conclus\~{o}es}

Embora 1900 seja considerado atualmente o ano do nascimento da f\'{\i}sica qu%
\^{a}ntica, a id\'{e}ia revolucion\'{a}ria dos quanta de energia n\~{a}o
despertou nenhuma aten\c{c}\~{a}o nos quatro anos seguintes. Foi apenas em
1905, com o trabalho de Einstein introduzindo a hip\'{o}tese dos quanta de
luz (f\'{o}tons) que o conceito de Planck come\c{c}ou a ser reconhecido. V%
\'{a}rias explica\c{c}\~{o}es foram aventadas. Klein\cite{klein62} argumenta
que a teoria da radia\c{c}\~{a}o n\~{a}o era o centro das aten\c{c}\~{o}es
em f\'{\i}sica na \'{e}poca, tendo em vista as grandes descobertas na virada
do s\'{e}culo: raios-X (1895), radioatividade (1896), el\'{e}tron (1897),
dentre outras. Al\'{e}m disto, um conjunto de eminentes cientistas,
liderados por Wilhelm Ostwald (1853-1932), atacava com furor os fundamentos
da teoria cin\'{e}tica. E em sua teoria, como vimos, embora Planck n\~{a}o
tenha usado o m\'{e}todo de Boltzmann (a distribui\c{c}\~{a}o mais prov\'{a}%
vel \'{e} aquela que maximiza a entropia do sistema), se baseia na rela\c{c}%
\~{a}o fundamental entre a entropia e probabilidade.

Apesar da maioria dos historiadores reconhecer Planck como o fundador da
teoria qu\^{a}ntica, h\'{a} pelo menos uma exce\c{c}\~{a}o ilustre. Thomas
Kuhn \ (1923-1976), o grande fil\'{o}sofo e historiador da ci\^{e}ncia,
sustenta\cite{kuhn78} que o conceito da descontinuidade qu\^{a}ntica nasceu
nos trabalhos de Einstein, seguido de Hendrik Lorentz (1853-1928) e
Ehrenfest entre os anos de 1906-1908 e n\~{a}o no trabalho de Planck, embora
obviamente este fosse uma importante contribui\c{c}\~{a}o. Para Kuhn, o
racioc\'{\i}nio de Planck foi completamente cl\'{a}ssico: ``embora a
estrutura do cont\'{\i}nuo de energia seja determinado pelo elemento de
energia $h$, o movimento dos osciladores de Planck permanece cont\'{\i}%
nuo...e nenhum dos trabalhos publicados, manuscritos conhecidos, ou
fragmentos autobiogr\'{a}ficos sugere que a id\'{e}ia de restringir as
energias dos ressonadores a um conjunto discreto de valores lhe ocorreu at%
\'{e} que outros o for\c{c}aram a reconhecer durante 1906 e nos anos
seguintes''. Realmente, nas suas \textit{Lectures} sobre a radia\c{c}\~{a}o
do calor\cite{planck06}, durante o semestre de ver\~{a}o de 1906-97 na
Universidade de Berlim, em que apresenta em detalhes a sua teoria, n\~{a}o h%
\'{a} men\c{c}\~{a}o \`{a} descontinuidade, nenhuma f\'{o}rmula como $%
u=nh\nu $. A \'{u}nica discretiza\c{c}\~{a}o aparece ao computar a
probabilidade de uma distribui\c{c}\~{a}o de energia, como fizera em seu
trabalho de 1900.

Apenas em outubro de 1908, numa carta a Lorentz, Planck referiu-se \`{a}
quantiza\c{c}\~{a}o de energia e a necessidade de uma descontinuidade:%
\footnote{%
Cita\c{c}\~{a}o da Ref. \cite{kuhn84}, p. 238.}

\begin{quotation}
{\small [A excita\c{c}\~{a}o dos ressonadores] n\~{a}o corresponde \`{a}
conhecida lei do p\^{e}ndulo simples; pelo contr\'{a}rio, existe um certo
limiar; o ressonador n\~{a}o responde a todas excita\c{c}\~{o}es muito
pequenas; e se responde \`{a}s maiores, o faz somente de modo que sua
energia seja um m\'{u}ltiplo inteiro do elemento de energia }$h\nu ${\small %
, tal que o valor instant\^{a}neo da energia \'{e} sempre representado por
tal m\'{u}ltiplo inteiro.}

{\small Em suma, eu poderia dizer que fa\c{c}o duas hip\'{o}teses:}

{\small - A energia do ressonador em um dado instante \'{e} }$gh\nu ${\small %
\ (}$g${\small \ um n\'{u}mero inteiro ou }$0${\small );}

{\small - A energia emitida e absorvida por um ressonador durante um
intervalo de tempo contendo bilh\~{o}es de oscila\c{c}\~{o}es (e portanto
tamb\'{e}m a energia m\'{e}dia de um oscilador) \'{e} a mesma que a equa\c{c}%
\~{a}o do p\^{e}ndulo.}
\end{quotation}

Teria sido Planck mais um ``son\^{a}mbulo'' da ci\^{e}ncia na provocativa
concep\c{c}\~{a}o\cite{koestler89} de Arthur Koestler (1905-1983)? Uma
discuss\~{a}o desta controv\'{e}rsia est\'{a} fora do escopo deste artigo
por falta de espa\c{c}o e compet\^{e}ncia do autor e sugiro as Refs. \cite%
{isis79}, \cite{galison81} e \cite{kuhn84} ao leitor interessado.

Para concluir, gostaria de salientar uma das verifica\c{c}\~{o}es
experimentais mais marcantes e precisas da lei de Planck e retornar aos
versos de Caetano. O Universo est\'{a} repleto de uma radia\c{c}\~{a}o c\'{o}%
smica de fundo a uma temperatura de $2,73$K, que \'{e} a mais importante evid%
\^{e}ncia da teoria do \textit{big bang,} apoiada na expans\~{a}o e
resfriamento do Universo com o tempo. Esta radia\c{c}\~{a}o \'{e} o mais
antigo f\'{o}ssil referente a um per\'{\i}odo em que a mat\'{e}ria (pr\'{o}%
tons e el\'{e}trons) estava em equil\'{\i}brio t\'{e}rmico com a radia\c{c}%
\~{a}o eletromagn\'{e}tica com todas as freq\"{u}\^{e}ncias. Quando o
Universo se esfriou a $T=3000$K -- a mat\'{e}ria j\'{a} era constitu\'{\i}da
de hidrog\^{e}nio at\^{o}mico --, a intera\c{c}\~{a}o com a radia\c{c}\~{a}o
de dava apenas nas freq\"{u}\^{e}ncias das respectivas linhas espectrais do
hidrog\^{e}nio. Nesta \'{e}poca, a maior parte da radia\c{c}\~{a}o se
separou da mat\'{e}ria, esfriando-se, a entropia constante, at\'{e} a atual
temperatura de $2,73$K.

\FRAME{fhFU}{2.8193in}{2.6498in}{0pt}{\Qcb{Distribui\c{c}\~{a}o espectral da
radia\c{c}\~{a}o c\'{o}smica de fundo correspondente \`{a} radia\c{c}\~{a}o
de um corpo negro a temperatura $T=2,73$ K.}}{\Qlb{Fig4}}{cobe.tif}{\special%
{language "Scientific Word";type "GRAPHIC";maintain-aspect-ratio
TRUE;display "USEDEF";valid_file "F";width 2.8193in;height 2.6498in;depth
0pt;original-width 7.6925in;original-height 7.2272in;cropleft "0";croptop
"1";cropright "1";cropbottom "0";filename 'cobe.tif';file-properties
"XNPEU";}}

A primeira evid\^{e}ncia da radia\c{c}\~{a}o f\'{o}ssil foi encontrada por
Arno Penzias (1933-) e Robert Wilson (1936-) em 1964. Um l\'{u}cido e
atraente relato da hist\'{o}ria desta descoberta e sua explica\c{c}\~{a}o 
\'{e} dado na Ref.\cite{weinberg80}. A distribui\c{c}\~{a}o espectral da
radia\c{c}\~{a}o de fundo, as \textit{microondas c\'{o}smicas}, foi obtida a
partir dos anos 90 pela miss\~{a}o Cosmic Background Explorer (COBE).\cite%
{cobe90} A Fig. \ref{Fig4} mostra a intensidade espectral como fun\c{c}\~{a}%
o da freq\"{u}\^{e}ncia, com intensidade m\'{a}xima na regi\~{a}o de
microondas. Os desvios da lei de Planck s\~{a}o m\'{\i}nimos (algumas partes
por milh\~{a}o) e s\~{a}o devidos a flutua\c{c}\~{o}es primordiais que
levaram ao aparecimento das gal\'{a}xias.

\section{Dados Biogr\'{a}ficos}

Planck nasceu em Kiel, Alemanha, no dia 23 de abril de 1858, filho de um
professor de direito constitucional da universidade local, Julius Wilhelm, e
de Emma (n\'{e}e Patzig) Planck. Seu pai vinha de uma fam\'{\i}lia de acad%
\^{e}micos (pai e av\^{o} eram professores de teologia da Universidade de G%
\"{o}ttingen). Recebeu a educa\c{c}\~{a}o inicial em Kiel e Munique e
estudou f\'{\i}sica e matem\'{a}tica nas Universidades de Munique
(1874-1877) e Berlim (1877-1888). Foi aluno de Hermann von Helmholtz
(1821-1894) e Gustav Kirchhoff em Berlim. Obteve o doutorado \textit{summa
cum laude} da Universidade de Munique, em 1879, com uma tese sobre a concep%
\c{c}\~{a}o de entropia no trabalho de Rudolf Clausius da segunda lei da
termodin\^{a}mica e no ano seguinte tornou-se \textit{Privatdozent} em
Munique. Em 1885, foi para a Universidade de Kiel, como \textit{%
Extraordinariat} (professor associado) e quatro anos depois sucedeu
Kirchhoff na Universidade de Berlim, tendo assumido a c\'{a}tedra de f\'{\i}%
sica te\'{o}rica em 1892 e permanecendo a\'{\i} at\'{e} a sua aposentadoria
em 1926. Foi membro da Academia Prussiana de F\'{\i}sica, desde 1894, e foi
eleito membro da \textit{Royal Society} de Londres em 1926. Foi agraciado
com o Pr\^{e}mio Nobel de F\'{\i}sica, em 1918,\ ``por seu trabalho sobre o
estabelecimento e desenvolvimento da teoria dos quanta elementares''.

Sua carreira cient\'{\i}fica inicial foi devotada ao estudo da segunda lei
da termodin\^{a}mica, especialmente o conceito de entropia com aplica\c{c}%
\~{o}es ao problema de equil\'{\i}bro f\'{\i}sico e qu\'{\i}mico, como transi%
\c{c}\~{a}o de fases e dissocia\c{c}\~{a}o eletrol\'{\i}tica. Foi
profundamente influenciado por Clausius e sempre muito preocupado em defini%
\c{c}\~{o}es claras dos conceitos fundamentais. Muito embora n\~{a}o
simpatizasse muito, naquela \'{e}poca, com o trabalho de Boltzmann, ficou ao
seu lado na famosa disputa contra Ostwald e os partid\'{a}rios do \
`Energismo'. Por volta de 1894, desviou a sua aten\c{c}\~{a}o para um novo
campo de estudos: a radia\c{c}\~{a}o do calor. Poss\'{\i}veis raz\~{o}es
foram a sua cren\c{c}a na import\^{a}ncia dos argumentos termodin\^{a}micos
no eletromagnetismo e o interesse geral nos fen\^{o}menos das ondas
eletromagn\'{e}ticos provocado pelas bem sucedidas experi\^{e}ncias de
Heinrich Hertz (1857-1894).

Planck publicou 235 trabalhos sobre ci\^{e}ncia e filosofia, conforme consta
da lista da Academia Prussiana de Ci\^{e}ncias, incluindo muitos discursos
relacionados com as suas fun\c{c}\~{o}es como ``Secret\'{a}rio Perp\'{e}%
tuo'' da Academia Prussiana (que n\~{a}o tinha o cargo de presidente), no per%
\'{\i}odo de 1918-1938, e confer\^{e}ncias sobre assuntos gerais. Em f\'{\i}%
sica, os principais campos foram Termodin\^{a}mica (primeiro e continuado
amor), Teoria Qu\^{a}ntica (a partir de 1900), e Teoria da Relatividade
Especial (principalmente no per\'{\i}odo 1906-1908).\footnote{%
Foi Planck quem corrigiu o \^{e}rro de Einstein em definir $\mathbf{F}=m%
\mathbf{a}$, e como conseq\"{u}\^{e}ncia, obtendo, em sua teoria relativ%
\'{\i}stica, diferentes massa transversal e longitudinal do el\'{e}tron.
Usando $\mathbf{F}=dp/dt,$ Planck mostrou que uma boa defini\c{c}\~{a}o de
momentum seria $\mathbf{p}=m_{0}\mathbf{v}/\sqrt{1-v^{2}/c^{2}\text{ }}$ e
portanto uma massa isotr\'{o}pica dependente da velocidade.\cite{miller81}}

\FRAME{fhFU}{2.0738in}{2.6507in}{0pt}{\Qcb{Max Planck }}{\Qlb{Fig5}}{%
fotoplanck.gif}{\special{language "Scientific Word";type
"GRAPHIC";maintain-aspect-ratio TRUE;display "USEDEF";valid_file "F";width
2.0738in;height 2.6507in;depth 0pt;original-width 3.9003in;original-height
5.0004in;cropleft "0";croptop "1";cropright "1";cropbottom "0";filename
'fotoplanck.gif';file-properties "XNPEU";}}

A Fig. \ref{Fig5} \'{e} o retrato de um Planck j\'{a} maduro, que desfrutou
de enorme prest\'{\i}gio na Alemanha e junto \`{a} comunidade cient\'{\i}%
fica, n\~{a}o apenas pela import\^{a}ncia de seu trabalho, mas tamb\'{e}m
por suas qualidades pessoais: car\'{a}ter, integridade moral, patriotismo e
lideran\c{c}a. Era muito benquisto por seus colegas e alunos. Quando
rejeitou o convite da Universidade de Viena para suceder Boltzmann, optando
por permanecer em Berlim, seus alunos festejaram alegremente com uma
passeata de tochas. Segundo v\'{a}rios historiadores, seu conservadorismo em
f\'{\i}sica o tornou relutante em aceitar o conte\'{u}do revolucion\'{a}rio
de sua pr\'{o}pria teoria qu\^{a}ntica. Durante anos, tentou ajustar o seu
conceito de quantum dentro da estrutura da f\'{\i}sica cl\'{a}ssica. Foi
presidente da Sociedade Kaiser Wilhelm para a Promo\c{c}\~{a}o da Ci\^{e}%
ncia de 1930 a 1937, reassumindo em 1945-56, um per\'{\i}odo particularmente
muito dif\'{\i}cil para a ci\^{e}ncia alem\~{a}.

Durante o nazismo, Planck permaneceu na Alemanha, mas se op\^{o}s a algumas
pol\'{\i}ticas governamentais, tendo tentado em v\~{a}o, em 1933, dissuadir
Hitler da expuls\~{a}o de cientistas judeus das universidades alem\~{a}s.
Criticou ainda a proposta de cientistas alem\~{a}es, liderados por Philipp
Lenard (1862-1957) e Johannes Stark (1874-1957) da cria\c{c}\~{a}o de uma `f%
\'{\i}sica ariana'. Um depoimento importante sobre as posi\c{c}\~{o}es de
Planck nesta \'{e}poca foi dado Werner Heisenberg (1901-1976).\cite%
{heisenberg} Fez uma defesa apaixonada de Einstein nos tempos duros da
repress\~{a}o. Em um discurso na Academia, em 11 de maio de 1933, afirmou:
``Eu acredito exprimir a opini\~{a}o de meus colegas de Academia e da
maioria dos f\'{\i}sicos alem\~{a}es ao dizer que o Sr. Einstein n\~{a}o 
\'{e} apenas um dos f\'{\i}sicos fora de s\'{e}rie, mas o Sr. Einstein \'{e}
o f\'{\i}sico, cujos trabalhos publicados em nosso s\'{e}culo, atingiram a
profundidade e import\^{a}ncia que somente pode ser igualada \`{a}s realiza%
\c{c}\~{o}es de Johannes Kepler e Isaac Newton''.\footnote{%
Cita\c{c}\~{a}o na Ref. \cite{broda83}.}

Planck apreciava m\'{u}sica tendo sido um bom pianista -- chegou a pensar em
seguir uma carreira profissional. Casou-se com Marie Merck, falecida em
1909, e em segundas n\'{u}pcias com sua prima, Marga von H\"{o}sslin. Planck
sofreu muito com as duas guerras. Na primeira, um de seus filhos, de um
total de cinco, morreu e outro foi enforcado pela Gestapo, em 1945, acusado
de participa\c{c}\~{a}o em um compl\^{o} para matar Hitler.\footnote{%
Alguns historiadores sugerem uma vingan\c{c}a pessoal de Hitler pela sua
defesa de Einstein e contra
\par
a expuls\~{a}o dos cientistas judeus.}

Einstein apreciava muito o trabalho e a pessoa de Planck. Em seu obitu\'{a}%
rio (1948), Einstein escreveu:\cite{einstein94}

\begin{quotation}
{\small Um homem a quem foi dado aben\c{c}oar o mundo com uma grande id\'{e}%
ia criativa n\~{a}o precisa do louvor, da posteridade. Sua pr\'{o}pria fa%
\c{c}anha j\'{a} lhe conferiu uma d\'{a}diva maior....Foi a lei da radia\c{c}%
\~{a}o de Planck, que forneceu a primeira demonstra\c{c}\~{a}o rigorosa --
independente de outras suposi\c{c}\~{o}es -- das magnitudes absolutas dos 
\'{a}tomos. Mais que isso, ele mostrou convincentemente que, al\'{e}m da
estrutura at\^{o}mica da mat\'{e}ria, h\'{a} uma esp\'{e}cie de estrutura at%
\^{o}mica da energia, regida pela constante universal }$h,$ {\small que
Planck introduziu}$.${\small \ [A sua descoberta] abalou toda a estrutura da
mec\^{a}nica e da eletrodin\^{a}mica cl\'{a}ssicas e imp\^{o}s \`{a} ci\^{e}%
ncia uma nova miss\~{a}o: a de encontrar uma nova base conceitual para toda
a f\'{\i}sica}.
\end{quotation}

\textit{Agradecimentos}: Sou muito grato ao Prof. Arthur Miller, do \textit{%
London College,} por disponibilizar, h\'{a} tempos, refer\^{e}ncias bibliogr%
\'{a}ficas relevantes e por conversas sobre o ensino da hist\'{o}ria da f%
\'{\i}sica. Agrade\c{c}o ainda aos Prof. Guilherme F. Leal Ferreira por
discuss\~{o}es sobre o artigo e o Prof. Salomon Mizrahi pela leitura atenta
do manuscrito.

\end{document}